\documentclass[12pt,english,preprint]{aastex}
\usepackage{mathptmx}

\usepackage[T1]{fontenc}
\usepackage[latin9]{inputenc}
\usepackage{graphicx}

\makeatletter

\usepackage{graphicx}

\def\mathcal{{\it }}

\def\ln{{\rm ln}}

\usepackage{hyperref}
\doublespace

\makeatother

\usepackage{babel}
\begin{document}

\title{Magnetic and density spikes in cosmic ray shock precursors }

\author{M.A. Malkov$^{1}$, R.Z. Sagdeev$^{2}$ and P.H. Diamond$^{1,3}$}

\affil{$^{1}$CASS and Department of Physics, University of California,
San Diego, La Jolla, CA 92093\newline$^{2}$University of Maryland,
College Park, Maryland 20742-3280, USA\newline$^{3}$WCI Center for
Fusion Theory, National Fusion Research Institute, Gwahangno 113,
Yuseong-gu, Daejeon 305-333, Republic of Korea}

\email{mmalkov@ucsd.edu}
\begin{abstract}
In shock precursors populated by accelerated cosmic rays (CR), the
CR return current instability is believed to significantly enhance
the pre-shock perturbations of magnetic field. We have obtained \emph{fully-nonlinear}
exact ideal MHD solutions supported by the CR return current. The
solutions occur as localized spikes of circularly polarized Alfven
envelopes (solitons, or breathers). As the conventional (undriven)
solitons, the obtained magnetic spikes propagate at a speed $C$ proportional
to their amplitude, $C=C_{A}B_{{\rm max}}/\sqrt{2}B_{0}$. The sufficiently
strong solitons run thus ahead of the main shock and stand in the
precursor, being supported by the return current. This property of
the nonlinear solutions is strikingly different from the linear theory
that predicts non-propagating (that is, convected downstream) circularly
polarized waves. The nonlinear solutions may come either in isolated
pulses (solitons) or in soliton-trains (cnoidal waves). The morphological
similarity of such quasi-periodic soliton chains with recently observed
X-ray stripes in Tycho supernova remnant (SNR) is briefly discussed.
The magnetic field amplification determined by the suggested saturation
process is obtained as a function of decreasing SNR blast wave velocity
during its evolution from the ejecta-dominated to the Sedov-Taylor
stage.
\end{abstract}


\section{Introduction\label{sec:Introduction}}

The nonresonant cosmic ray (CR) return current instability (also termed
as Bell's instability) is expected to bootstrap the acceleration of
CR in shocks by enhancing the magnetic field in the shock precursor.
The most unstable is a circularly polarized, field aligned, aperiodic
mode, similar to the internal kink (Kruskal-Shafranov) mode in plasmas
(see, e.g., \citealp{RyutovKink06}). In the context of the CR acceleration
in shocks, it was also studied by \citet{Acht83FireHose,ShapiroQuest98GeoRL}. 

\citet{Bell04} reawakened the interest in this instability by emphasizing
its role in magnetic field amplification and suggested its saturation
due to magnetic tension. But, since the growth rate decreases with
the wave number only as $\sqrt{k}$, the magnetic tension term does
not stabilize the long waves. This opens the door for a strong, $\delta B\gg B_{0}$
field amplification. The caveat is that the non-propagating long waves
have limited (precursor-crossing) time to grow. By contrast, the nonlinear
solutions, that we present in this paper, can stand off in the flow
ahead of the shock, thus warranting the saturation. 

There have been considerable efforts to understand the Bell's mode
saturation mechanisms, with a strong emphasis on the MHD and PIC simulations
\citep{Bell04,Pelletier06,Vladimirov06,Reville07,Niemiec08,ZirakBell08,BykovBell09,MelroseBell09,SpitkBell09,StromanNiem09,Dieckmann10}.
As first demonstrated in 3D MHD simulations by \citet{Bell05} (see
also \citealp{Niemiec08}), the saturation is achieved when the Ampere
force expels plasma and the helical magnetic field radially, thus
forming plasma cavities.  The instability appears to saturate only
in 3D, or at least requires a quasi-2D dynamics, perpendicular to
the ambient field. However, the fastest growing modes are field-aligned,
i.e., at least initially one-dimensional. Therefore, it is necessary
to understand structures that form at the 1D phase and particularly
the nonlinear mechanisms of their saturation and propagation ahead
of the shock. These structures may in the main cease to grow before
the subsequent 3D dynamics kick in largely by spreading the saturated
turbulence energy in $k$-space. Although this scenario may appear
to be at odds with many simulations, recent Chandra observations of
the Tycho supernova remnant (SNR), for example, indicate the presence
of quasi-1D structures (stripes), inconsistent with the quasi-isotropic
nonlinear dynamics observed in those simulations \citep{Eriksen11}.
Moreover, while being very useful for our understanding of CR instabilities,
simulations cover only a tiny fraction of the dynamical range of typical
SNR-shock acceleration process and introduce artificial dissipation
in essentially collisionless plasmas. 

Alfven waves usually saturate by modulational instability. However,
being a strong MHD \emph{aperiodic }instability, Bell's instability
hampers direct applications of standard methods, such as the weak-turbulence
theory \citep{SagdGal69}. The latter typically deals with propagating
and weakly interacting eigen modes and, as a driver amplifies them,
they cascade the wave energy to the dissipation scale. The Bell's
linear mode does not propagate (in the linear approximation), and
does not even exists without the driving current. The lack of long
wave stabilization is also based on the comparison of linear contributions
to the square of the growth rate of the driving current ($\propto k$)
and magnetic tension ($\propto k^{2},$ eq.{[}\ref{eq:omegalin}{]}
below). A clue to saturation in a similar system of the pressure-anisotropy-driven
fire-hose instability is provided by an exact solution due to \citet{BerezinSagdeev69}.
While at peaks of magnetic energy it takes nearly all the instability
free energy ($B_{\perp}^{2}/8\pi\sim P_{\parallel}-P_{\perp}\gg B_{0}^{2}/8\pi$),
on the average only the moderate field amplification $B_{\perp}\sim B_{0}$
is observed. 

In this paper we present an exact solution of the \emph{current-driven}
MHD equations (e.g., CR return current). It differs from the linearly
growing solution in that it propagates with the velocity proportional
to its (constant) amplitude and is spatially localized.

\section{Basic equations \label{sec:BasicEquations} }

The linear theory of Bell's instability indicates that the fastest
growing modes are directed along the ambient magnetic field \citep{Bell05}.
Therefore, we will consider 1D magnetohydrodynamic (MHD) equations
in a CR shock precursor using a coordinate system with the axis $x$
along the field. The general 1D equations read:

\begin{equation}
\frac{d\rho}{dt}+\rho\frac{\partial}{\partial x}U_{x}=0,\label{eq:cont}
\end{equation}

\begin{equation}
\frac{dU_{x}}{dt}=-\frac{1}{\rho}\frac{\partial}{\partial x}\frac{B_{\perp}^{2}}{8\pi},\label{eq:parmom}
\end{equation}

\begin{equation}
\frac{d\mathbf{U}_{\perp}}{dt}=\frac{B_{0}}{4\pi\rho}\frac{\partial\mathbf{B_{\perp}}}{\partial x}+\frac{1}{c\rho}\mathbf{J}\times\mathbf{B_{\mathbf{\perp}}},\label{eq:perpmom}
\end{equation}

\begin{equation}
\frac{d\mathbf{B}_{\perp}}{dt}=B_{0}\frac{\partial\mathbf{U}_{\perp}}{\partial x}-\mathbf{B}_{\perp}\frac{\partial U_{x}}{\partial x}.\label{eq:induc}
\end{equation}
Here $d/dt\equiv\partial/\partial t+U_{x}\partial/\partial x$, $\rho$
is the gas density, $U_{x}$,$B_{0}$ and $\mathbf{U_{\mathbf{\perp}}}$,$\mathbf{B}_{\perp}$
are the gas velocity and magnetic field components along the field
and in the ($y,z$)-plane, respectively. The x-component of magnetic
field $B_{x}=B_{0}=const$ because of $\nabla\cdot\mathbf{B=0}$.
In eq.(\ref{eq:perpmom}) we have included the plasma return current
by representing the total plasma current as $\mathbf{J}_{tot}=\left(c/4\pi\right)\nabla\times\mathbf{B}+\mathbf{J}$,
where the part of the plasma current $\mathbf{J=-}\mathbf{J}_{CR}$
compensates the CR current. Eq.(\ref{eq:perpmom}) implies that in
our reference frame $\mathbf{J}\times\mathbf{B}_{0}=0$. We neglect
the thermal and CR pressure, as \citet{Bell04} did. It should be
noted, however, that these CR pressure gradient drives an acoustic
instability of the shock precursor (also called Drury's instability,
\citealp{Drury84,Dorfi84}). Moreover, the acoustic instability grows
faster than the Bell's instability for $\beta=8\pi P/B_{0}^{2}<1$
(see \citealp{MDS10PPCF} where the studies of evolution, saturation,
as well as the associated particle transport and cascading of magnetic
energy, are also referenced).

Returning to eqs.(\ref{eq:cont}-\ref{eq:induc}), it is convenient
to introduce a Lagrangian mass coordinate $\xi$: 

\begin{equation}
d\xi=\frac{\rho}{\rho_{0}}\left(dx-U_{x}dt\right),\label{eq:Lagra}
\end{equation}
where $\rho_{0}$ is the background density. Considering scales shorter
than the precursor size, we treat $\rho_{0}$ and the bulk plasma
speed $U_{x0}$ as coordinate independent ($U_{x0}=0$ in the plasma
frame). 

Next, we reduce eqs.(\ref{eq:cont}-\ref{eq:induc}) to the following
system of two equations that describe the magnetic field and density
perturbations:

\begin{eqnarray}
\frac{\partial^{2}}{\partial t^{2}}\frac{B}{\rho}-C_{A}^{2}\frac{\partial^{2}}{\partial\xi^{2}}\frac{B}{\rho_{0}} & = & \frac{i}{c\rho_{0}}B_{0}J\frac{\partial}{\partial\xi}\frac{B}{\rho}\label{eq:b}\\
\frac{\partial^{2}}{\partial t^{2}}\frac{\rho_{0}^{2}}{\rho}+\frac{\partial^{2}}{\partial\xi^{2}}\frac{\left|B\right|^{2}}{8\pi} & = & 0,\label{eq:rho}
\end{eqnarray}
where 

\[
B=B_{y}+iB_{z}\;\;{\rm and}\;\; C_{A}^{2}=\frac{B_{0}^{2}}{4\pi\rho_{0}}.
\]
The r.h.s. of eqs.(\ref{eq:b}) is the instability driver. Without
it, the equations describe the conventional MHD modes, propagating
at an arbitrary angle to the ambient magnetic field. By choosing the
averaged components $\overline{B}_{y}=\overline{B}_{z}=0$, we restrict
our treatment to the parallel propagation along the x-direction.

\section{Traveling wave solutions\label{sec:Traveling-wave-solutions}}

We look for the solutions of the system given by eqs.(\ref{eq:b})
and (\ref{eq:rho}) in the form of a traveling wave: 

\begin{equation}
B=B_{{\rm max}}v\left(\zeta\right)e^{-i\omega t}\label{eq:Btrw}
\end{equation}

\[
\rho=\rho\left(\zeta\right)
\]
where $\zeta=\xi-Ct$, $C$ is the (constant) propagation speed of
the traveling wave, $B_{{\rm max}}$ is the wave amplitude that we
specify in eq.(\ref{eq:rhoInt}) below, and $\omega=\Re\omega$ is
the wave frequency. Note that for $\omega\neq0$, the solution is
not steady in any reference frame. For that reason, the spatially
localized version of this solution is some times called 'breather'
as opposed to the soliton, customary to $\omega=0$ case. Integrating
then eq.(\ref{eq:rho}) twice, we obtain

\begin{equation}
\frac{\rho_{0}}{\rho}=1-\frac{\left|B\right|^{2}}{B_{{\rm max}}^{2}}\label{eq:rhoInt}
\end{equation}
where $B_{{\rm max}}^{2}\equiv8\pi\rho_{0}C^{2}$. We have chosen
the integration constants in such a way that $B\to0$ for $\rho\to\rho_{0}$
(background plasma) and $B\to B_{{\rm max}}$ for $\rho\to\infty$
(flow stagnation point, if present). This sets the interval for variation
of $v\left(\zeta\right):\;0<v<1$.

Substituting $B$ from eq.(\ref{eq:Btrw}) and $\rho$ from the last
equation, eq.(\ref{eq:b}) yields

\begin{equation}
\frac{\partial^{2}}{\partial\zeta^{2}}\left(a-\left|v\right|^{2}\right)v-iK\frac{\partial}{\partial\zeta}\left(1-\left|v\right|^{2}\right)v-\frac{\omega^{2}}{C^{2}}\left(1-\left|v\right|^{2}\right)v=0.\label{eq:v}
\end{equation}
Here we have used the following notations

\begin{equation}
K=\frac{B_{0}J}{c\rho_{0}C^{2}}-2\frac{\omega}{C},\;\; a=1-2\frac{B_{0}^{2}}{B_{max}^{2}}=1-\frac{C_{A}^{2}}{C^{2}},\label{eq:Kanda}
\end{equation}
where $C_{A}^{2}=B_{0}^{2}/4\pi\rho_{0}$. The linear dispersion relation
can be recovered by letting $v\left(\zeta\right)\propto e^{ik\zeta}$,
$v\to0$ in eq.(\ref{eq:v}):

\begin{equation}
\omega=kC\pm\sqrt{k^{2}C_{A}^{2}+B_{0}Jk/c\rho_{0}}.\label{eq:omegalin}
\end{equation}
The arbitrary propagation speed $C$ is a parameter of a Galilean
transformation (zero in the plasma frame), while the imaginary part
of $\omega$ is an invariant of such transformation as it should be.
In the nonlinear treatment the wave velocity with respect to the plasma
depends on the wave amplitude (nonlinear dispersion relation). Meanwhile,
the linear instability occurs in the long wave limit for $B_{0}Jk<0$.
It should be emphasized that only if the quadratic $B$ term is neglected
in eq.(\ref{eq:rhoInt}), is there no coupling to the density modulations
in eq.(\ref{eq:b}). It is interesting to note that in the strong
nonlinear limit $B_{max}/B_{0}\to\infty$, eq.(\ref{eq:v}) degenerates
into a linear equation for the function $v\left(1-\left|v\right|^{2}\right)$.
This limit, however, cannot be understood without the nonlinear solution. 

To find such solution, we write

\begin{equation}
v\left(\zeta\right)=\sqrt{w}e^{i\Theta}\label{eq:vexprep}
\end{equation}
where $w\left(\zeta\right)\ge0$. Substituting $v$ from eq.(\ref{eq:vexprep})
into eq.(\ref{eq:v}) and separating the imaginary part, for the phase
$\Theta$ we obtain the following equation:

\begin{equation}
\frac{d\Theta}{ds}=\frac{wP\left(w\right)+A}{w(a-w)^{2}},\label{eq:thetaprime}
\end{equation}
where

\[
P\equiv w^{2}-(3a+1)w/2+a
\]
We have introduced a new variable $s=K\zeta/2$ and an integration
constant $A$. We may choose it by specifying the properties of the
solution sought. The regularity of $\Theta$ at $w=0$ implies $A=0$.
Next, taking the real part of eq.(\ref{eq:v}) and using eq.(\ref{eq:thetaprime})
with $A=0$, for $w\left(s\right)$ we obtain

\begin{equation}
\frac{d^{2}f}{ds^{2}}-\frac{w^{2}P^{2}}{f^{3}}+2\sqrt{w}\left(1-w\right)\left[\frac{wP}{f^{2}}-2\frac{\omega^{2}}{C^{2}K^{2}}\right]=0\label{eq:f}
\end{equation}
where we have denoted

\[
f\left(w\right)\equiv\sqrt{w}(a-w)
\]
Eq.(\ref{eq:f}) can be readily integrated by multiplying it by $df/ds$.
We choose the integration constant to select an isolated pulse (soliton)
solution of eq.(\ref{eq:f}), i.e. $w\to0,$ as $s\to\pm\infty$.
Then, the first integral reads

\begin{equation}
\left(\frac{dw}{ds}\right)^{2}-\frac{w^{2}}{\left(3w-a\right)^{2}\left(a-w\right)^{2}}\sum_{n=0}^{4}C_{n}w^{n}=0\label{eq:wfulleq}
\end{equation}
where 

\[
\begin{array}{cc}
C_{0}= & 4a^{2}\left(a\mu^{2}-1\right)\\
C_{1}= & 2a\left[2\left(3+a\right)-a\mu^{2}\left(7+a\right)\right]\\
C_{2}= & 8\mu^{2}a\left(a+2\right)-a^{2}-14a-9\\
C_{3}= & 2\left[2\left(3+a\right)-\mu^{2}\left(3+5a\right)\right]\\
C_{4}= & 4\left(\mu^{2}-1\right)
\end{array}
\]
with

\[
\mu^{2}\equiv4\omega^{2}/K^{2}C^{2}=\left(1-\frac{B_{0}J}{2c\rho_{0}C\omega}\right)^{-2}.
\]
A useful analogy between nonlinear waves and nonlinear oscillators
(e.g., \citealp{Sagdeev66}) suggests to interpret the first term
eq.(\ref{eq:wfulleq}) as kinetic and the second term as potential
energy. The 'oscillator's coordinate' $w>0$, as a function of 'time'
$s$, leaves $w=0+$ at $s=-\infty$ and returns there at $s=+\infty$.
\emph{Periodic} solutions (cnoidal waves) can also be easily by changing
the integration constant.

The amplitude $w_{0}\left(a,\mu\right)$ of the localized solution
(soliton) is obviously determined by the smallest positive root of
the polynomial in eq.(\ref{eq:wfulleq}), so that the {}``oscillator''
bounces between $w=0$ and $w=w_{0}$. In the simplest case of a small
amplitude solution 

\[
w_{0}\approx-C_{0}/C_{1}\ll a<1
\]
(where $C_{0}>0$ and $C_{1}<0$) the solution has a classical soliton
profile

\begin{equation}
w\left(s\right)=\frac{w_{0}}{\cosh^{2}\left(\frac{\sqrt{C_{0}}}{2a^{2}}s\right)}\label{eq:cosh}
\end{equation}
Apart from the condition $a\mu^{2}>1$ (to ensure $w_{0}>0$), i.e.

\[
1-\sqrt{1-C_{A}^{2}/C^{2}}<\frac{B_{0}J}{2c\rho_{0}\omega C}<1+\sqrt{1-C_{A}^{2}/C^{2}}
\]
and the technical restriction $a\mu^{2}-1\ll1$ (to neglect the $n>1$
terms in eq.{[}\ref{eq:wfulleq}{]}), this solution imposes no further
constraints on $\omega$ and $C$. However, it has a very strong amplitude
limitation, $a\mu^{2}-1\ll1$ (virtually a wave packet of linear waves).
We are interested in an opposite case of highly superalfvenic solitons
that are not convected rapidly with the flow into the sub-shock and
may stay ahead of it, when $C\gg C_{A}$. 

A relation between $\omega$ and $C$ (nonlinear dispersion relation)
arises from the extension of the above solution to larger $w_{0}$.
Clearly, we have to pass the point $w=a/3$ smoothly which requires
a double root of the polynomial in eq.(\ref{eq:wfulleq}) at $w=a/3$:

\[
\sum_{n=0}^{4}C_{n}\left(\frac{a}{3}\right)^{n}=\sum_{n=1}^{4}nC_{n}\left(\frac{a}{3}\right)^{n-1}=0.
\]
Interestingly, the both conditions are met simultaneously as soon
as the following dispersion relation is satisfied:

\begin{equation}
a\mu^{2}=\frac{9-a}{8}\label{eq:finaldisprel}
\end{equation}
Recalling that the small amplitude soliton $w_{0}\ll1$ branches off
from the trivial solution at the threshold $a\mu^{2}=1$, in the case
of $C_{A}\ll C$, i.e. $a\approx1$, we may accept eq.(\ref{eq:finaldisprel})
to be valid in the entire parameter range $a\mu^{2}>1$. 

Let us rewrite the above dispersion relation as follows

\[
\omega=\frac{k_{J}C}{M_{A}^{2}\left(1\pm\sqrt{\left(1-M_{A}^{-2}\right)/\left(1+1/8M_{A}^{2}\right)}\right)},
\]
where we have defined the linear instability wave number (see eq.{[}\ref{eq:omegalin}{]})
as $k_{J}=2\pi J/cB_{0}$. Strong solitons with $M_{A}\equiv C/C_{A}\gg1$
have either high or low frequency: $\omega=\left(16/9\right)k_{J}C$,
$\omega=k_{J}C/2M_{A}^{2}$. The spatial scale of the solitons, given
by the 'wave number' $K$, eq.(\ref{eq:Kanda}), can be represented
as follows, Fig.\ref{fig:Dispersion}:

\begin{equation}
K=\pm\frac{2k_{J}}{M_{A}^{2}}\frac{\sqrt{\left(1-M_{A}^{-2}\right)/\left(1+1/8M_{A}^{2}\right)}}{1\pm\sqrt{\left(1-M_{A}^{-2}\right)/\left(1+1/8M_{A}^{2}\right)}}\label{eq:kappaOfM}
\end{equation}
It is interesting to note that both solutions disappear (spread to
infinity) in the limit $J\to0$, although they have disparate scales,
particularly for $M_{A}\gg1$. Therefore, the external current is
essential and there is no transition to conventional simple wave MHD
solutions for the vanishing CR-current.

\begin{figure}
\includegraphics[bb=20bp 200bp 792bp 612bp,scale=0.7]{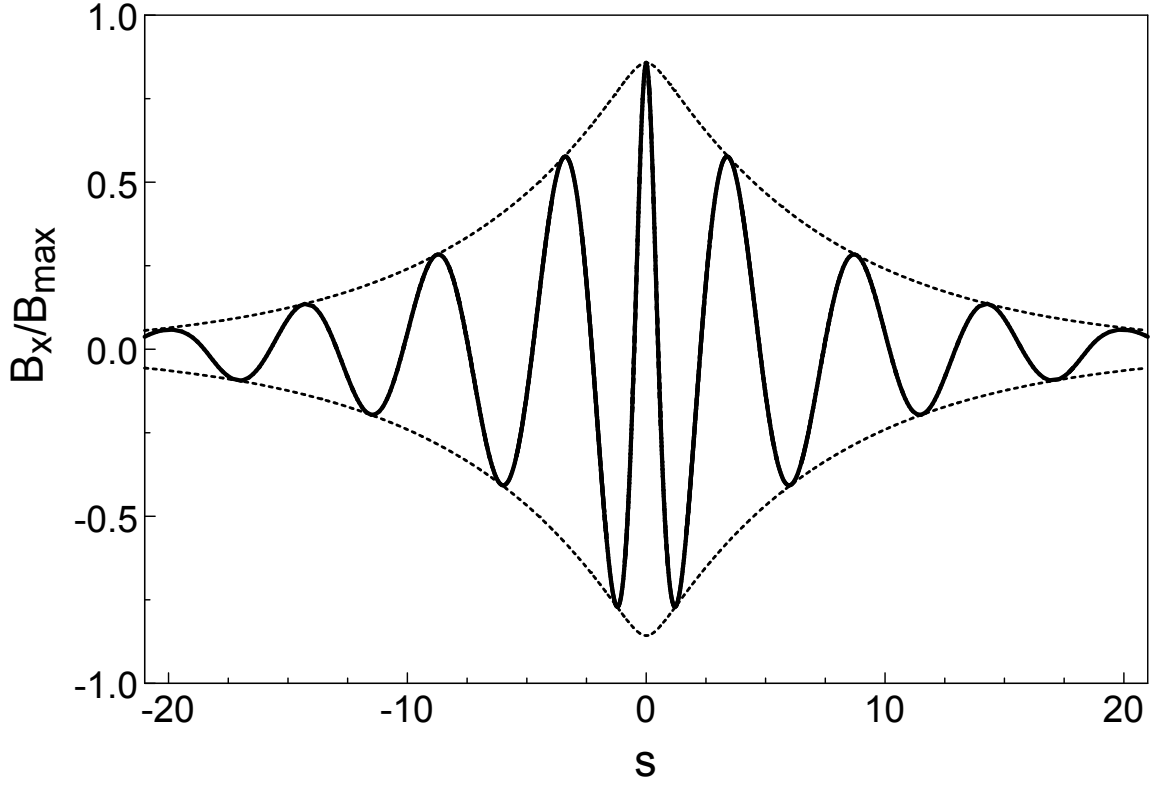}

\caption{$B_{x}$-component of the soliton solution as a function of dimensionless
coordinate $s=K\zeta/2$ in units of $B_{{\rm max}}$ ($B_{{\rm max}}^{2}\equiv8\pi\rho_{0}C^{2}$):
$B_{x}/B_{{\rm max}}=\sqrt{w}\cos\left(\Theta\right)$ shown with
the solid line and the amplitude envelope, $\pm\sqrt{w}$ (dashed
line). The soliton Mach number $M_{A}=C/C_{A}=3$ corresponds to the
amplitude parameter $a=1-M_{A}^{-2}$=8/9. \label{fig:Bx}}
\end{figure}

\begin{figure}
\includegraphics[bb=20bp 150bp 792bp 612bp,scale=0.7]{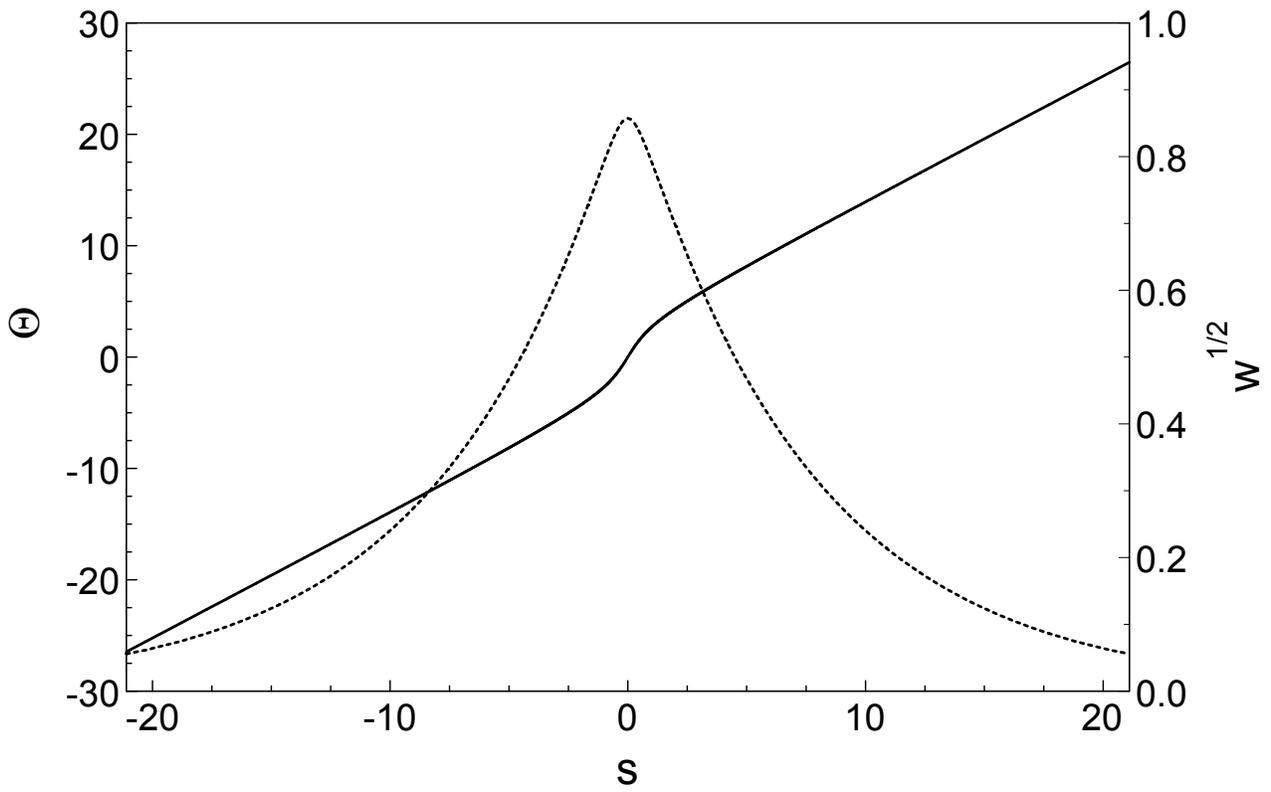}

\caption{Phase $\Theta$ as a function of coordinate $s$, shown with the solid
line. The amplitude of the soliton is shown with the dashed line.
\label{fig:Phase}}
\end{figure}

To obtain the spatial structure of the above solutions, we substitute
eq.(\ref{eq:finaldisprel}) into eq.(\ref{eq:wfulleq}). The latter
takes the following simple form

\begin{equation}
\left(\frac{dw}{ds}\right)^{2}=\frac{1-a}{2a}\frac{w^{2}}{\left(a-w\right)^{2}}Q^{2}\left(w\right)\label{eq:wfineq}
\end{equation}
where

\[
Q^{2}\equiv w^{2}-2hw+a;\;\;\; h=(a+3)/4.
\]
Eq.(\ref{eq:wfineq}) can be reduced to a quadrature:

\begin{equation}
s\left(w\right)=\sqrt{\frac{2a}{1-a}}\left[\cosh^{-1}\frac{h-w}{\sqrt{h^{2}-a}}+\sqrt{a}R\right],\label{eq:sOFw}
\end{equation}
where

\[
R=\ln\frac{\sqrt{a}+w-Q}{\sqrt{a}-w+Q}-\ln\frac{\sqrt{a}-\sqrt{h^{2}-a}+h}{\sqrt{a}+\sqrt{h^{2}-a}-h}.
\]
Using eqs.(\ref{eq:thetaprime}) and (\ref{eq:wfineq}), the solution
for the phase $\Theta\left(w\right)$ can be reduced to another quadrature

\[
\Theta=\sqrt{\frac{2a}{1-a}}\left[\cosh^{-1}\frac{h-w}{\sqrt{h^{2}-a}}+R/\sqrt{a}\right]-
\]

\begin{equation}
2\left[\tan^{-1}\left(\frac{Q-w+a}{\sqrt{a\left(1-a\right)/2}}\right)-\tan^{-1}\left(\frac{\sqrt{h^{2}-a}-h+a}{\sqrt{a\left(1-a\right)/2}}\right)\right]\label{eq:ThetaOFw}
\end{equation}
The $B_{x}$-component of the solitary solution is shown in Fig.\ref{fig:Bx}
(the $e^{-i\omega t}$ -factor omitted). The wave packet in the compressed
area becomes more oscillatory, as may also be seen from Fig.\ref{fig:Phase},
which shows the soliton phase $\Theta$ as a function of dimensionless
coordinate $s$. The local dimensionless wave number stays approximately
constant ($d\Theta/ds\approx1$), apart from the above phase steepening
near the maximum amplitude, where $d\Theta/ds\approx2$.

\section{Maximum Magnetic Field\label{sec:Maximum-Magnetic-Field}}

The isolated solitons obtained in this paper belong to a one parameter
family; the amplitude $B_{{\rm max}}$ or Mach number $M_{A}=B_{{\rm max}}/\sqrt{2}B_{0}$
can be used as such parameter. In a CR shock precursor the soliton
scale  is determined by the scale of seed waves for the subsequent
\emph{nonresonant }instability. The seed waves are \emph{resonantly
}excited upstream of the strong CR current zone by the high energy
CRs. Then, $K\sim r_{g}^{-1}(p_{*})$, where $r_{g}$ is the gyroradius
of the seed generating CRs of momentum $p=p_{*}$. This amounts to
$M_{A}^{2}=k_{J}r_{g}\left(p_{*}\right)$ for the upper (long-wave)
soliton branch in eq.(\ref{eq:kappaOfM}). Note that $p_{*}$ may
be $\ll p_{{\rm max}}$ due to a poor CR confinement in the range
$p_{*}<p<p_{{\rm max}}$ \citep{MD06}. If the CR current is sufficiently
strong, $k_{J}r_{g}\left(p_{*}\right)\gg1$ and only the upper-sign
soliton in eq.(\ref{eq:kappaOfM}) and Fig.\ref{fig:Dispersion} can
accommodate the requirement $K\sim r_{g}^{-1}(p_{*})$ for $M_{A}\gg1$.
Then, the maximum magnetic field for a given soliton can be written
as $B_{{\rm max}}^{2}/B_{0}^{2}=2M_{A}^{2}\approx2k_{J}r_{g}\left(p_{*}\right)$,
or 

\begin{equation}
B_{{\rm max}}^{2}=4\pi V_{{\rm s}}n_{CR}p_{*},\label{eq:BmaxSQ}
\end{equation}
where $n_{{\rm CR}}$ is the CR density. 

\begin{figure}
\includegraphics[bb=60bp 100bp 720bp 540bp,scale=0.9]{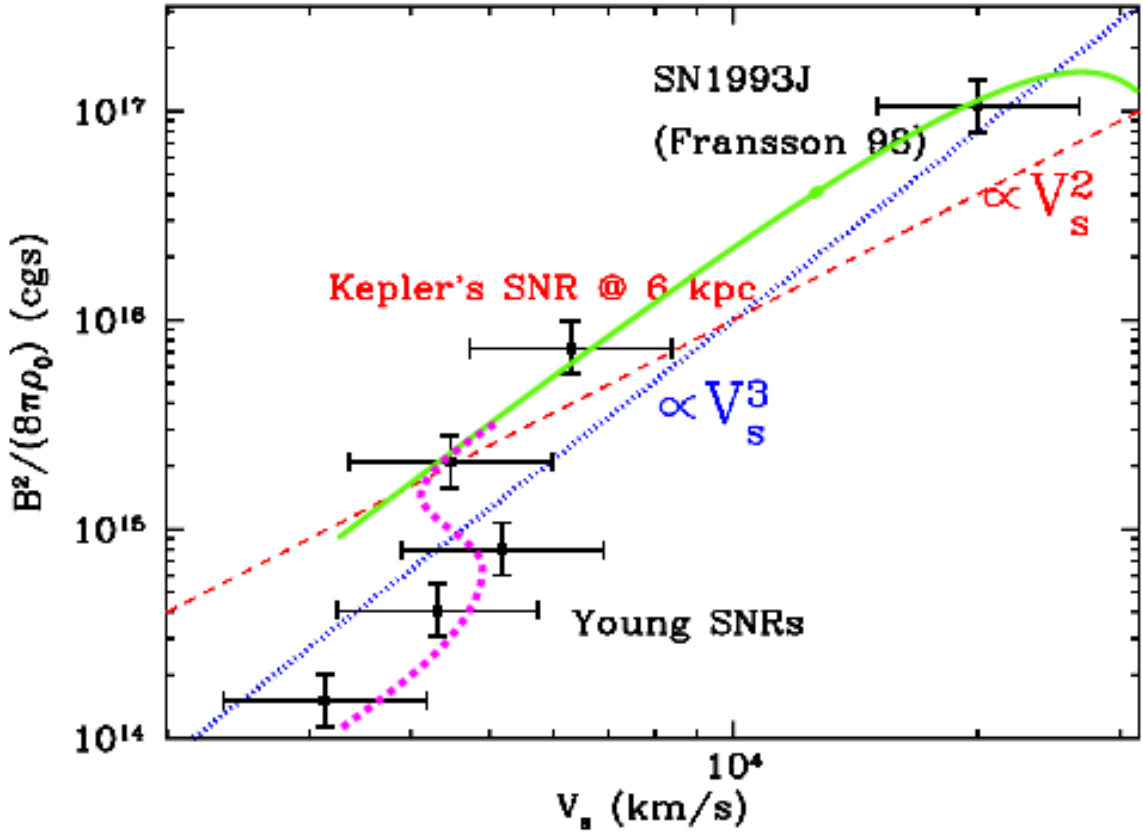}

\caption{The ratio of magnetic field energy to ambient density $\rho_{0}$
as a function of shock velocity $V_{{\rm s}}$ adopted from \citep{Vink08AIP}
(points with error bars). The blue dotted line is the scaling from
\citep{Bell04}, while the red dashed line is that of \citep{Vlk05},
both also taken from the Vink's compilation. An example of calculations,
described in Sec.\ref{sec:Maximum-Magnetic-Field} is shown with the
green line. The field energy declines beyond $V_{{\rm s}}\approx2.7\cdot10^{4}$
to vanish at $V_{0}\approx9\cdot10^{4}{\rm km/s}$ (not shown in the
plot). At lower $V_{{\rm s}}$, where the shock acceleration is inefficient,
this dependence breaks down and should transition to a low-efficiency
acceleration regime in a bistable fashion \citep{Drury81,m97b} (schematically
shown with the magenta dotted line). \label{fig:VinkFig} }
\end{figure}

The scaling of CR-enhanced magnetic energy with the ambient density
$\rho$ and shock velocity $V_{{\rm s}}$ is debated in the literature.
\citet{Bell04} suggested $B^{2}/\rho\propto V_{{\rm s}}^{3}$, whereas
\citet{Vlk05} indicate that $B^{2}/\rho\propto V_{{\rm s}}^{2}$.
The difference between the two scalings is whether a fraction of \emph{mechanical
energy flux }or \emph{momentum flux} goes into magnetic energy. By
contrast, eq.(\ref{eq:BmaxSQ}) constitutes the conversion of \emph{CR
energy} \emph{flux} into magnetic energy. \citet{Vink08AIP} summarizes
the information about the magnetic field from a number of SNR, with
the two phenomenological scalings superimposed, Fig.\ref{fig:VinkFig}.
Note that $B_{{\rm max}}$ in eq.(\ref{eq:BmaxSQ}) coincides with
the condition of magnetization (trapping by the wave) of the current-carrying
particles $k_{J}\left(B_{{\rm max}}\right)r_{g}\left(p_{*},B_{{\rm max}}\right)=1/2$,
which is also (formally) similar to the Bell's phenomenological condition
of balancing the Ampere force and the magnetic tension for the instability
saturation. However, the saturation mechanism behind eq.(\ref{eq:BmaxSQ})
is different in that $B_{{\rm max}}$ is only the peak magnetic field.
The magnetic energy density would be smaller by a soliton filling
factor $f_{{\rm s}}$ \citep[cf. ][]{BerezinSagdeev69}. More importantly,
the efficiency of CR acceleration and subsequent conversion of their
energy into magnetic field should depend on $J,\; V_{{\rm s}}$ and
other acceleration parameters which almost certainly rules out the
single power-law relation between $B^{2}/\rho$ and $V_{{\rm s}}$.

Therefore, we obtain such relation in a different way, which we outline
below and will describe in detail elsewhere. Consider a nominal SNR
with the shock speed $V_{{\rm s}}$ slowing down from an initial $V_{{\rm s}}=V_{0}=1.34\times V_{{\rm ST}}$,
to $V_{{\rm {\rm s}}}\approx0.1\times V_{0}$ (i.e., well into the
Sedov-Taylor phase) where $V_{{\rm ST}}=10400\times E_{51}^{1/2}\left(M_{{\rm e}}/M_{\odot}\right)^{1/2}km/s$
\citep{McKeeTruelove95}. Here $E_{51}$ is the explosion energy in
$10^{51}ergs$ and $M_{{\rm e}}$ -the ejecta mass. During its evolution,
the SNR should follow the points sampled from a set of supposedly
similar remnants in Fig.\ref{fig:VinkFig}. Using the $V_{{\rm s}}\left(t\right)$
dependence from \citep{McKeeTruelove95}, the momentum $p_{*}\sim p_{{\rm max}}\left(t\right)$
in the nonlinear acceleration regime from (\citealp{MDru01}, eq.{[}7.45{]}),
we express $p_{*}$ in eq.(\ref{eq:BmaxSQ}) as a function of $V_{{\rm s}}$.
Next, we obtain $n_{{\rm CR}}$ from eq.(15) in \citep{m97b} for
the evolving subshock strength with the particle injection rate held
approximately constant in the efficient acceleration regime (see eq.(37)
in the same reference). Using these results, we finally obtain from
eq.(\ref{eq:BmaxSQ}) an expression for $B_{{\rm max}}^{2}/\rho$,
again, as a function of $V_{{\rm s}}$. A preliminary example of such
calculation is shown in Fig.\ref{fig:VinkFig} with the green line.
In an intermediate range of $V_{{\rm s}}$ the scaling is $\propto B_{{\rm m}}^{2}/\rho\propto V_{{\rm s}}^{11/4}$
(close to the Bell's scaling) but it rolls over to turn to zero at
$V_{{\rm s}}=V_{0}\approx9\cdot10^{4}{\rm km/s}$. This is because
the magnetic field generation is pinned to the CRs {[}eq.(\ref{eq:BmaxSQ}){]}
which are not yet there at $V_{{\rm s}}=V_{0}$, i.e. at $t=0$. The
other strong deviation from a power-law should occur at lower $V_{{\rm s}}$
where acceleration bifurcates into a inefficient (test particle regime)
through a characteristic S-curve.

\section{Discussion\label{sec:Discussion}}

The purpose of this paper was to understand the nonlinear evolution
of the non-resonant current driven instability by studying saturated
nonlinear waves (solitons) as their ensemble (or that of their shock
counterparts, if dissipation is efficient) may comprise the asymptotic
state of the system. Such scenario is supported by simpler (but fully
integrable, e.g., \citealp{KaupNewell78}) weakly nonlinear MHD models,
such as the derivative nonlinear Schroedinger equation (DNLS, see
also \citealp{MjolhHada06} for a review). In such models, arbitrary
initial conditions evolve into an asymptotic state of quasi-independently
propagating solitons, very much similar to those found in the present
paper. The difference, however is that our system is driven by the
CR return current and its solutions do not transition into the MHD
solutions.

\begin{figure}
\includegraphics[bb=20bp 0bp 792bp 612bp,scale=0.7,angle=270]{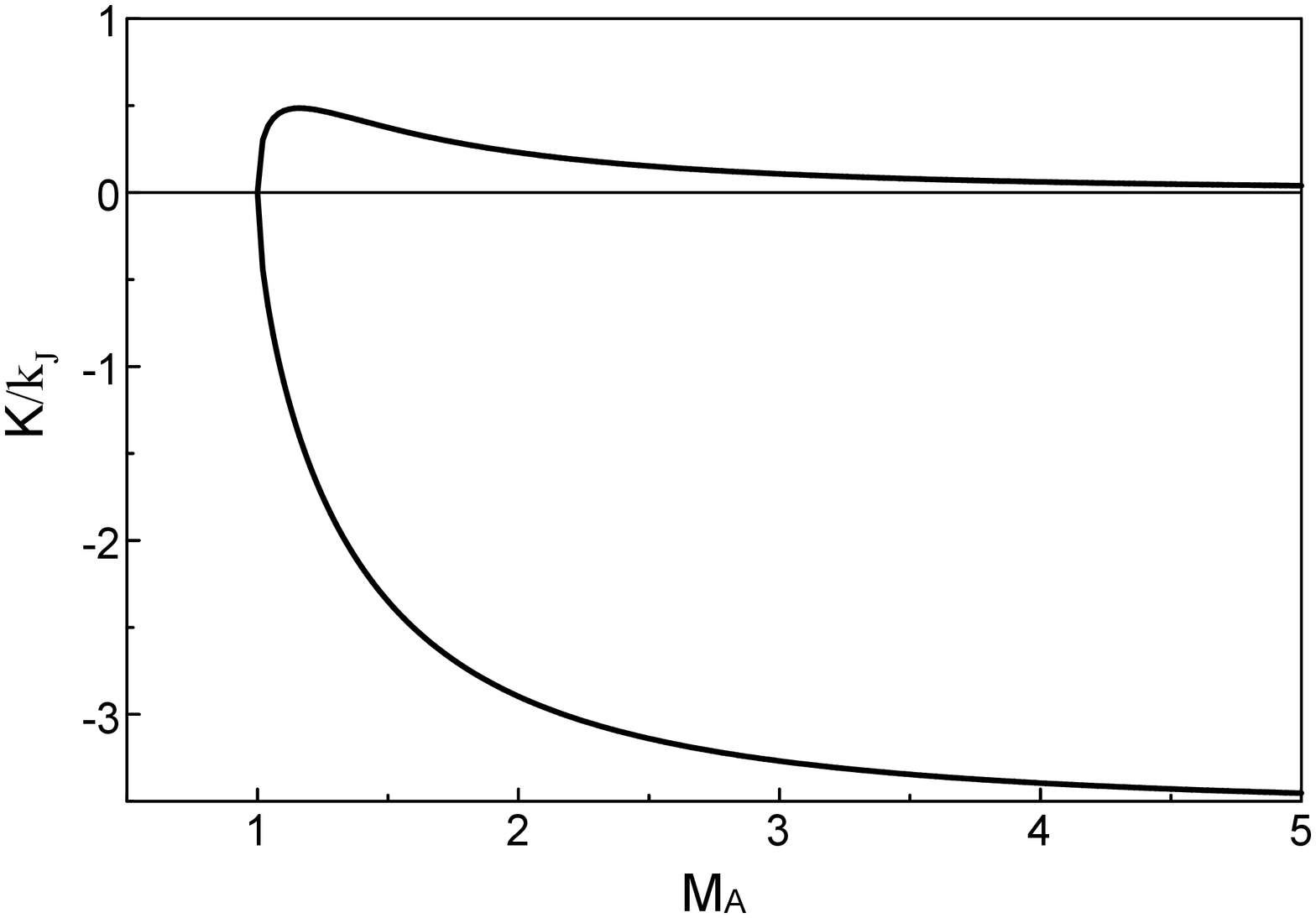}

\caption{Dispersive properties of the two types of solitons: the short scale
(lower branch) and the long scale (upper brunch). The soliton wave
number $K$ is shown in the units of $k_{J}=2\pi J/cB_{0}$.\label{fig:Dispersion}}
\end{figure}

The relevant question of soliton stability should be addressed in
2-3D setting. The 2-3D instability of a 1D soliton could comprise
a wave front self-focusing \citep{PassotSulem03} and thus elucidate
the subsequent 3D structures. Such studies are beyond the scope of
this letter, but a qualitative stability examination is in order.
It may be based on the nonlinear dispersion relation given by eq.(\ref{eq:kappaOfM})
and Fig.\ref{fig:Dispersion}. The parts of the dispersion curves
with $\partial\left|K\right|/\partial C<0$ (where $K$ and $C\propto M_{A}$
are the wave number and propagation speed) correspond to the solitons
with negative dispersion and should be stable. The oft-used justification
of stability is that a nonlinear steepening of the soliton's leading
edge generates higher wave number modes and they should not run faster
than the soliton itself (negative dispersion is thus required). It
should be also noted here that, once the soliton solutions of the
driven system are obtained, they can also be arranged in a quasi-periodic
or even chaotic soliton lattice. By adding weak dissipation, the leading
edges of these solitons can be converted into shock fronts \citep{Sagdeev66}
which usually increases the dissipation of the driver energy, thus
reducing the saturation amplitude.

To conclude, there exists upper bound on $B_{{\rm max}}$ since solitons
with $C/C_{A}=B_{{\rm max}}/\sqrt{2}B_{0}>V_{{\rm s}}/C_{{\rm A}}$
outrun the shock and cannot be sustained by the return current. However,
as \emph{transients,} they may promote particle acceleration far upstream
to synergistically supply themselves with the CR return current. This
might be a plausible scenario for much-discussed CR acceleration bootstrap
(e.g., \citealp{MDru01,BlandfordFunk07}). Furthermore, strong solitons
running ahead of the shock may become visible in X-rays as quasi-periodic
stripes, similar to those recently observed by Chandra \citep{Eriksen11}
in some parts of the Tycho SNR. The Eriksen's identification of the
stripe spacing with the maximum gyroradius of accelerated particles
is consistent with our determination of the soliton spacing in Sec.\ref{sec:Maximum-Magnetic-Field},
but with a lower than 2 PeV energy. The scale is set by the highest
energy particles ahead of the field amplification zone. The soliton
wave length should be noticeably shorter than the distance between
them. At the same time, similar structures may result from the nonlinear
evolution of the CR-pressure-driven acoustic instability studied earlier
by \citet{MD09}. Both the Drury's \citep{MDS10PPCF} and Bells's
instabilities (after adding dissipation to the soliton solution) should
result in shock-like nonlinear structures considerably shorter than
the conventional CR precursor of the standard Bohm diffusion model.
The magnetic field enhancement is clearly weaker in the acoustic case,
as it is merely due to the individual shock compression in the instability
generated shock-train. Besides, the soliton scenario is exciting as
it introduces these fascinating and ubiquitous objects (e.g., \citealp{AblowitzSegur81})
to the SNR physics. However, the dominant instability should be selected
on a case by case basis by treating the alternatives in a specific
shock environment.

\acknowledgements{}

Support by the Department of Energy, Grant No. DE-FG02-04ER54738 is
gratefully acknowledged.

\pagebreak

\end{document}